\documentclass[twocolumn,aps,showpacs,preprintnumbers,byrevtex,superscriptaddress]{revtex4}
\usepackage{graphicx}
\usepackage{dcolumn}
\usepackage{bm}
\begin{document}
\title{Quantum correlations in the soliton collisions}

\author{Ray-Kuang Lee$^{1,2}$, Yinchieh Lai$^2$, and Yuri S. Kivshar}

\affiliation{Nonlinear Physics Centre, Research School of Physical
Sciences and Engineering, The Australian National
University, Canberra, ACT 0200, Australia\\
$^2$Department of Photonics and Institute of Electro-Optical
Engineering, National Chiao-Tung University, Hsinchu 300, Taiwan}


\begin{abstract}
We study quantum correlations and quantum noise in the soliton collision described by a general two-soliton solution of the nonlinear Schr\"odinger equation, by using the back-propagation method.
Our results include the standard case of a $sech$-shaped initial pulse analyzed earlier.
We reveal that double-hump initial pulses can get more squeezed, and the squeezing ratio enhancement is due to the 
long collision period in which the pulses are more stationary.
These results offer promising possibilities of using higher-order solitons to generate strongly squeezed states for the quantum information process and quantum computation.
\end{abstract}

\maketitle

\section{Introduction}

Light squeezing is an important physical concept that continues to attract attention of researchers due to its potential for implementing quantum information processing and quantum computing.
As an alternative to single-photon schemes, the demonstration of the Einstein, Podolsky, and Rosen (EPR) paradox and quantum teleportation with
continuous variables have been realized experimentally by using the entanglement from squeezed states~\cite{Ou92, Furusawa98}. 
Moreover, experimental progress in the study of various quantum
information processing with squeezed states generated from optical
fibers has recently been
reported~\cite{Silberhorn01,Silberhorn02,Glockl03,Konig02}. To
increase the entanglement fidelity of continuous variables, the
enhancement of squeezing effect becomes very vital.

Original proposals to generate squeezed states from optical fibers
are based on the use of the fundamental solitons supported by the
Kerr nonlinearity of silica glass. Temporal (pulse) solitons in
optical fibers are described by the nonlinear Schr{\"{o}}dinger
equation (NLSE) that can exhibit quadrature-field squeezing
\cite{Carter87, Drummond87, Lai89a, Lai89b, Lai90} as well as
amplitude squeezing~\cite{Friberg, RK-fbg}, and both intra-pulse
and inter-pulse correlations~\cite{Schmidt00}. Besides the exact
quantum soliton NLSE solutions  constructed by using the Bethe
ansatz~\cite{Lai89b}, the quantum properties of temporal solitons
are well described by the linearization approach~\cite{Lai90} for
the average photon number as high as $10^9$. Based on this
linearization approach, many different numerical methods have been
developed during the past two decades in order to study quantum
noise associated with nonlinear pulse propagation, including the
positive-$P$ representation~\cite{Carter87, Drummond87},
back-propagation method~\cite{Lai95}, and the cumulant expansion
technique~\cite{Schmidt99}.

Experimentally, the soliton squeezing from a Sagnac fiber
interferometer, 1.7 dB below the shot noise, was first observed in
1991 by Rosenbluh and Shelby ~\cite{Rosenbluh}. Since that, larger
quadrature squeezing from fibers has been obtained with a
gigahertz Erbium-doped fiber lasers that allow to suppress the
guided acoustic-wave Brillouin scattering, and 6.1 dB noise
reduction below the shot noise has been reported~\cite{Yu01}. As
an attempt to enhance the soliton squeezing effect, one may
increase the energy of an optical soliton enhancing the importance
of nonlinear effects and, employing this idea, 7.1 dB
photon-number squeezing has been demonstrated by using spectral
filters~\cite{Werner97}.

However, it is known that the basic model of the pulse propagation
in optical fibers described by NLSE possesses more general
$N-$soliton solutions which can be obtained, for example, by
applying the inverse scattering transform~\cite{Zakharov}.
As was demonstrated, such higher-order ($N=2,3,\dots$) solitons can be
more squeezed since they contain $N^2$ times of the energy than
the fundamental soliton~\cite{Werner96,Yeang99,Schmidt00}; as an
example, up to 8.4 dB enhancement was predicted for the $N=2$
soliton states~\cite{Schmidt-oc}.

It is important to mention that all previous studies of the soliton quantum noise and quantum
squeezing of higher-order solitons have employed a very special
case of two-soliton states generated by a simple sech-like input
pulse. However, a general NLSE solution describing the $N$-soliton
state is characterized by N free parameters which can be
controlled independently. In this paper, we develop the theory of quantum noise and quantum
squeezing in the context of the multi-soliton states and apply it
to study squeezing of the general $N$-soliton bound states of
NLSE. In particular, we reveal that the conventional sech-like
single-hump pulses are not the most suitable pulses for generating
highly squeezed states and, using the case of the general
two-parameter solution for $N=2$ solitons as an example, we show
that an input double-hump soliton is better for generating
strongly squeezed states, even such solitons have the same energy
as the single-hump pulses.
The enhancement of the squeezing effect is explained by the long collision period of a double-hump soliton, consequently the pulse profile is more stationary for getting squeezed.
Since these double-hump solitons have also been generated in the fiber laser systems, more strongly squeezed state from optical fibers are expected to be realized with the current technology.

\section{Two-soliton bound states}

\begin{center}
\begin{figure}
\includegraphics[width=1.6in]{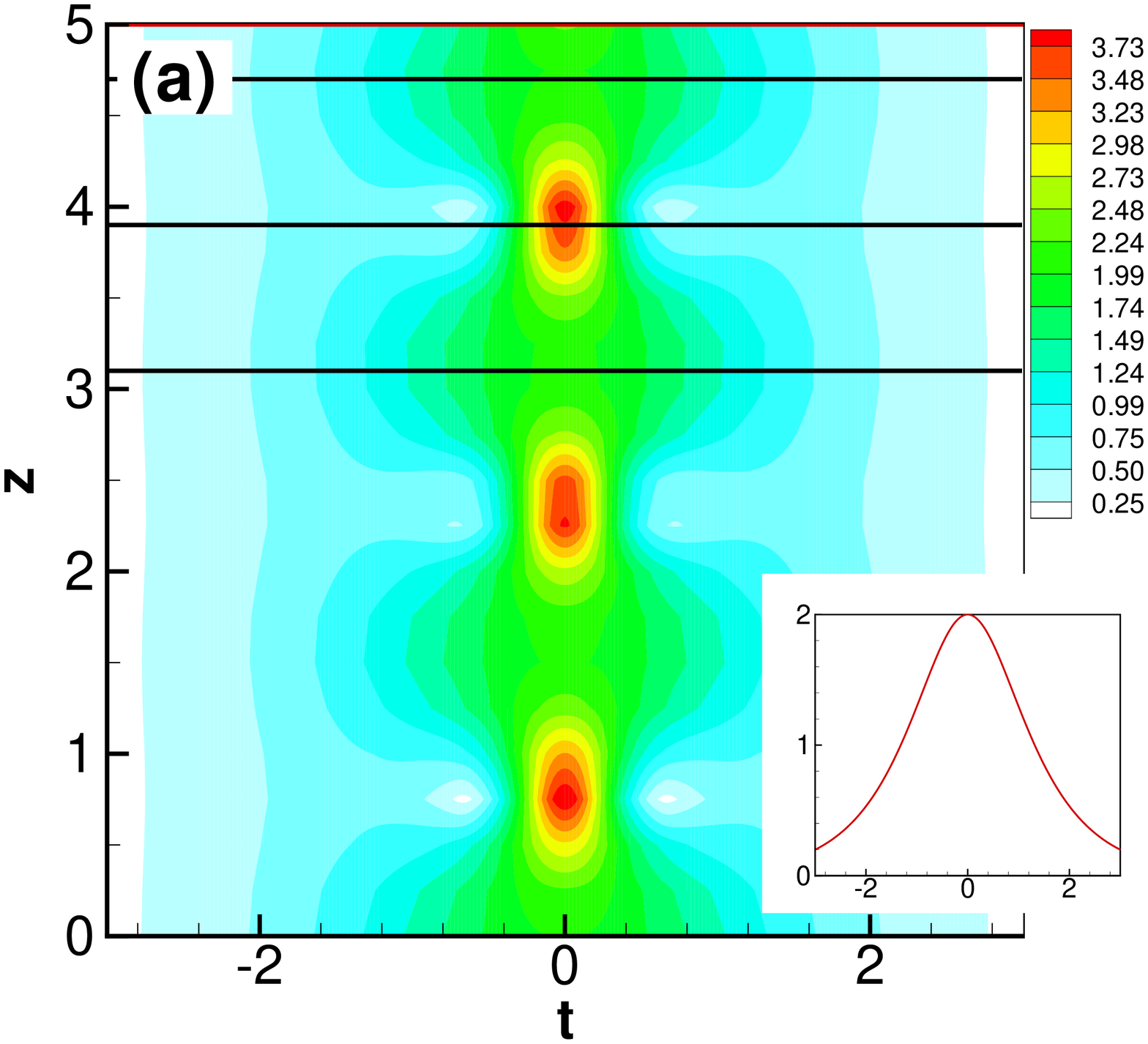}
\includegraphics[width=1.6in]{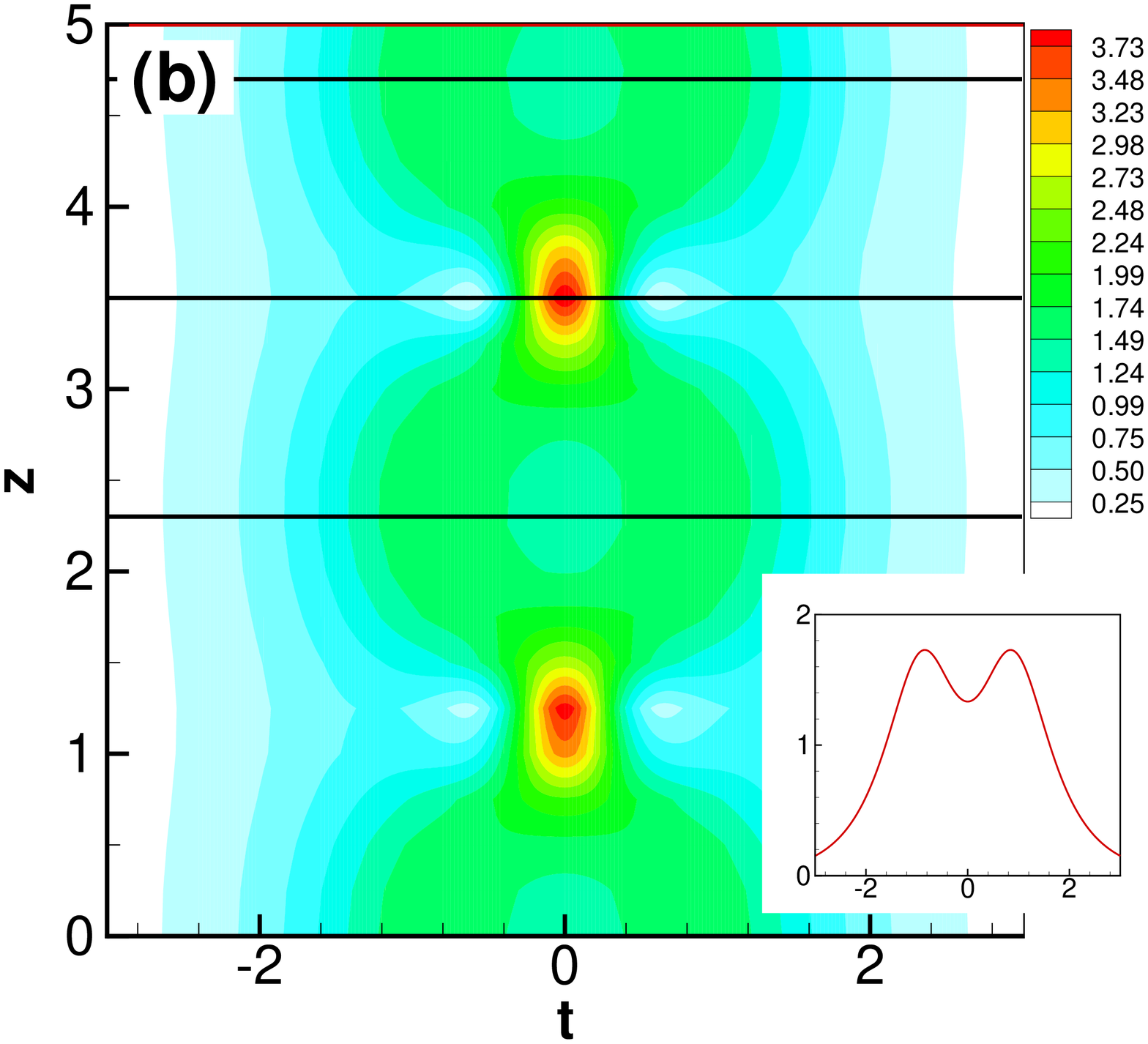}
\caption{The contour plots the evolution of the $N=2$ solitons with (a) $\eta_1:\eta_2=1:3$, and (b) $\eta_1:\eta_2=1:2$. The insets show the initial soliton profiles at $z = 0$. In all cases, $\eta_1+\eta=2$. The straight lines mark the propagation distances shown in Fig. \ref{fig_f3}.}
\label{fig_f1}
\end{figure}
\end{center}

To describe the pulses propagating in optical fibers with the
anomalous dispersion and Kerr-type nonlinearity, one employ the
NLSE model written for the normalized variables, $z$ and $t$,
\begin{equation}
i \frac{\partial U(z,t)}{\partial z}+\frac{\partial^2 U(z,t)}{\partial t^2}+|U(z,t)|^{2}U(z,t)=0,
\label{NLSE}
\end{equation}
where $U(z,t)$ is the pulse envelope.
According to the results of the inverse scattering transform \cite{Zakharov}, this equation
possesses exact solutions describing interaction of $N$ solitons,
which are characterized by a set of complex variables
$\{\lambda_j, C_j\}$, $j=1,2,\dots,N$. The complex $\lambda_j$ and
$C_j$ are the "poles" and the "residues" of the corresponding
scattering data \cite{Zakharov}.
In particular, such solutions describe the so-called $N-$soliton bound states (also called "breathers") when
all solitons have vanishing velocities at the infinity, and their
interaction leads to the formation of a spatially localized but
time-periodic state. The full set of such solutions for the $N=2$
soliton bound states can be written
as~\cite{s-laser,surface-mode},
\begin{eqnarray}
U(z,t) =4\eta_1
\frac{(\eta_1+\eta_2)}{|\eta_2-\eta_1|}\frac{A(z,t)}{B(z,t)}e^{2 i
\eta_1^2 z},
\end{eqnarray}
where
\begin{eqnarray}
A(z,t) &=& {\rm cosh}(2\eta_2 t)+\frac{\eta_2}{\eta_1}{\rm cosh}(2\eta_1 t)e^{2i(\eta_2^2-\eta_1^2)z},\\
B(z,t) &=& \frac{(\eta_1+\eta_2)^2}{(\eta_2-\eta_1)^2}{\rm
cosh}{[}2(\eta_2-\eta_1) z{]}\\\nonumber
&+& \frac{4\eta_1}{(\eta_2-\eta_1)^2}\cos{[}2(\eta_2^2-\eta_1^2) z{]} + {\rm cosh}{[}2(\eta_1+\eta_2) z{]}.
\end{eqnarray}
The two free parameters, $\eta_1$ and $\eta_2$, are the imaginary
parts of the poles in the scattering data, i.e. $\lambda_j=i \eta_j$, and the residues are related to the poles by the
relations
\[
C_j^2 = \frac{\prod_{k=1}^N (\eta_j+\eta_k)}{\prod_{k=1,k\neq j}^N |\eta_j-\eta_k|}.
\]
When $\eta_1:\eta_2 = 1:3$, the soliton solution at $z= 0$ has a specific,
sech-like single-hump initial profile,
\[
U(0,t) = \frac{2}{{\rm sech}(t)},
\]
as shown in the insert of Fig.~\ref{fig_f1}(a).
However, when the ratio of $\eta_1/\eta_2$ becomes larger than $1/3$, the initial
profile of the $N=2$ soliton solution becomes double-humped, as
shown in the insert of Fig.~\ref{fig_f1}(b) for the special case
of $\eta_1:\eta_2 = 1:2$. In spite of such a difference in the
soliton profiles, the soliton energy defined as
\[
P=\int|U(z,t)| d\,t^2,
\]
remains the same for the full set of the $N=2$ soliton solutions,
i.e. $P=8$ for $\eta_1+\eta_2 = 2$ and arbitrary ratio $\eta_1/\eta_2$.

\section{Soliton squeezing}

After knowing that the general solution for the $N=2$ solitons may have different initial profiles when we change the ratio $\eta_1/\eta_2$, we apply the \textit{back-propagation method} \cite{Lai95} to calculate the quantum fluctuations of a full set of the $N=2$ soliton solutions.
To evaluate the quantum fluctuations around the bound solitons, we replace the classical function $U(z,t)$ in Eq. (\ref{NLSE}) by the quantum-field operator variable, $\hat{U}(z,t)$, which satisfies the equal-coordinate bosonic commutation relations.
Next, we substitute the expansion $\hat{U}=U_{0}+\hat{u}$ into Eq. (\ref{NLSE}) to linearize it around the classical solution $U_{0}$ for the soliton containing a large number of photons. Then we calculate the quantum uncertainty of the output field by back-propagating the output field to the input field with the assumption that the statistics of the input quantum-field operators obey the Poisson distribution.
In particular, we calculate the squeezing ratio, defined below, of the output field based on the homodyne detection
scheme~\cite{Haus90,Lai93},
\begin{equation}
R(L)\equiv \frac{\mathrm{var}[\langle f_{L}(t)|\hat{u}(L,t)\rangle]}{\mathrm{var}[\langle f_{L}(t)|\hat{u}(0,t)\rangle]},
\nonumber
\end{equation}
where $\mathrm{var}[\cdot ]$ stands for the variance, $f_{L}(t)$
is the normalized classical pulse solution in the output with an
adjustable phase shift,
\[
f_{L}(t)=\frac{U_{0}(L,t)e^{i\theta }}{\sqrt{\int_{-\infty }^{+\infty}d\,t|U_{0}(L,t)|^{2}}},
\]
which acts as a local oscillator. The optimal (minimum) value of
the squeezing ratio $R(z)$ can be chosen by varying the parameter
$\theta$. When $\theta =0$, the in-phase quadrature component is
detected, and when $\theta =\pi /2$, the out-of-phase quadrature
component is detected.
\begin{center}
\begin{figure}
\includegraphics[width=3.0in]{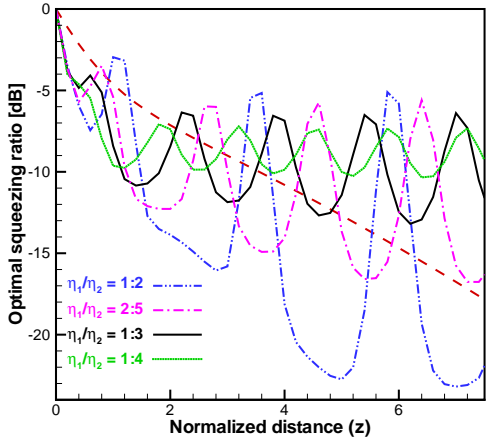}
\caption{Optimal squeezing ratio vs. propagation distance for the
$N=2$ solitons with different values of the ratio of $\eta_1/\eta_2$, for  $\eta_1+\eta_2=2$. Dashed line shows the
optimal squeezing ratio curve for the case of the $N=1$ soliton.}
\label{fig_f2}
\end{figure}
\end{center}

Based on the formulation above, now we calculate the optimal quadrature squeezing ratios for a full set of the $N=2$ soliton bound states in Fig.~\ref{fig_f2}.
Compared to the optimal squeezing ratio of the fundamental soliton (dashed line), all $N=2$ solitons get more squeezed in the beginning of their propagation because they contain more energy \cite{Yeang99, Schmidt00}.
Moreover, after a certain propagation distance, the optimal squeezing ratio of the $N=2$ soliton state change periodically due to the oscillating behavior of the breather, with the period of the $N=2$ soliton~\cite{s-laser},
\begin{eqnarray}
Z_p = \frac{\pi}{(\eta_2+\eta_1)(\eta_2-\eta_1)}.
\label{eq_p}
\end{eqnarray}
However, if we compare the optimal squeezing ratios between the $N=2$ solitons with different ratios $\eta_1/\eta_2$, we find that solitons with larger $\eta_1/\eta_2$ are more squeezed, even all of them have the same energy.

The reason that a $N=2$ soliton with a larger $\eta_1/\eta_2$ gets more squeezed can be inducted from the comparison of the optimal squeezing ratio with that of the $N=1$ soliton.
When the propagation distance is large enough (beyond the propagation range shown in the Fig. \ref{fig_f2}),  there is a oscillating tail in the optimal squeezing ratio of the $N=1$ soliton coming form the continuum part of the noise due to the use of the same pulse profile as the local oscillator.
But basically the optimal squeezing ratio of the fundamental soliton increases monochromatically along the propagation distance due to the stationary characteristic of the pulses.
On the contrary, the oscillation nature of $N=2$ solitons prevent the increase of the optimal squeezing ratio after a certain degree.
Since a $N=2$ soliton with larger ratio of $\eta_1/\eta_2$ has a longer collision period, as shown in Eq. (\ref{eq_p}), it is this longer collision period that makes the pulse to behave more stationary, and more squeezed.

\begin{center}
\begin{figure}
\includegraphics[width=1.6in]{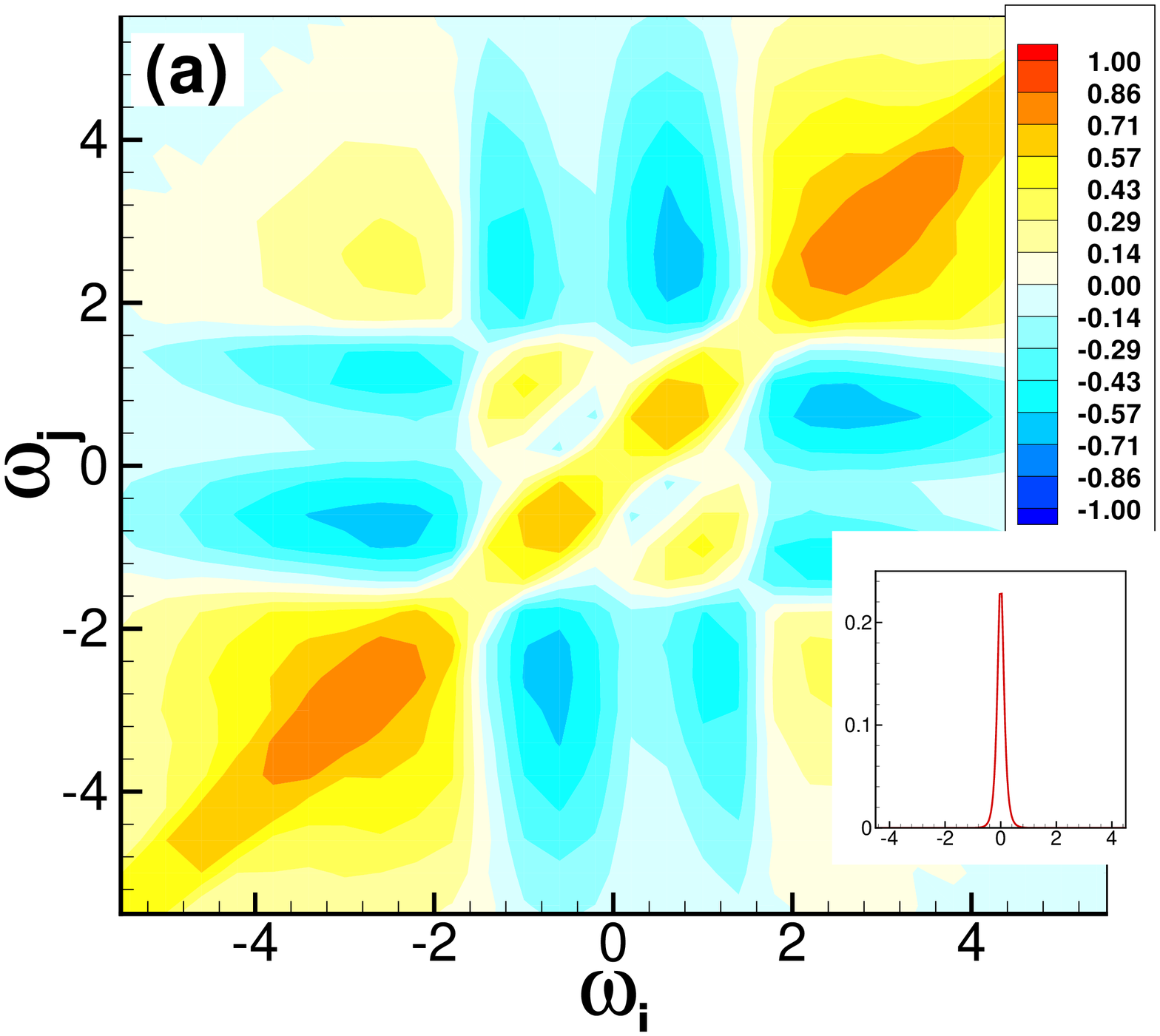}
\includegraphics[width=1.6in]{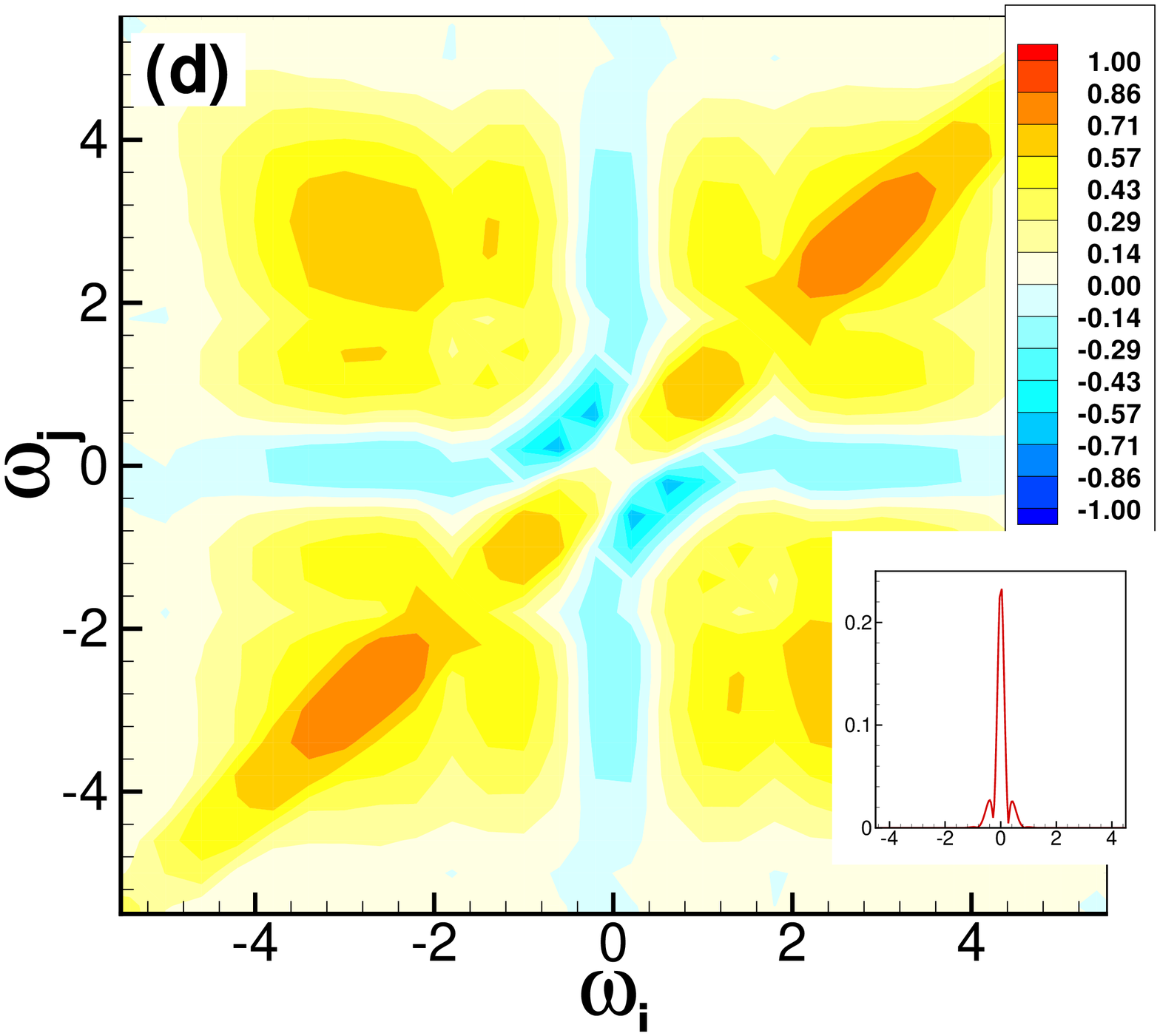}
\includegraphics[width=1.6in]{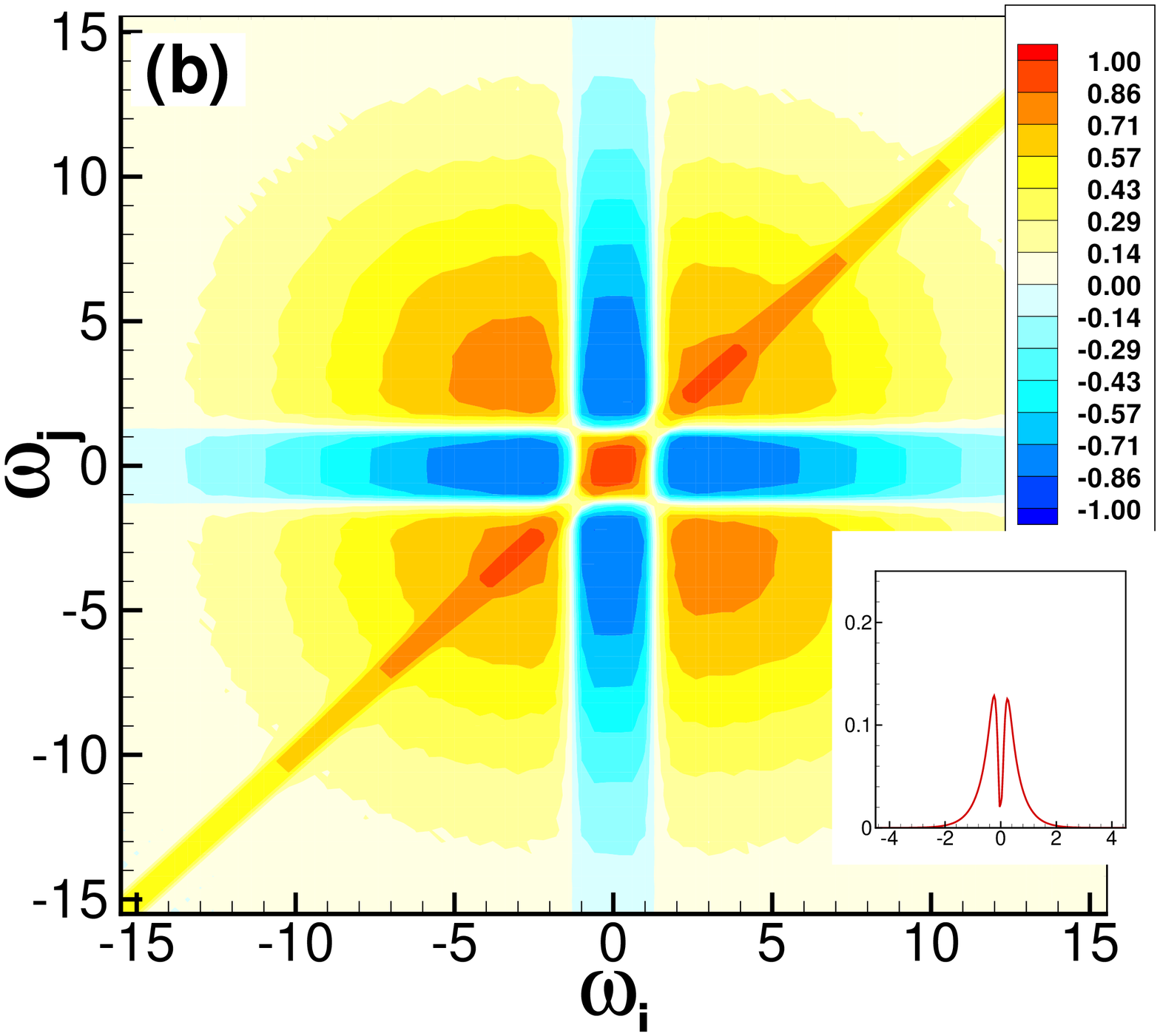}
\includegraphics[width=1.6in]{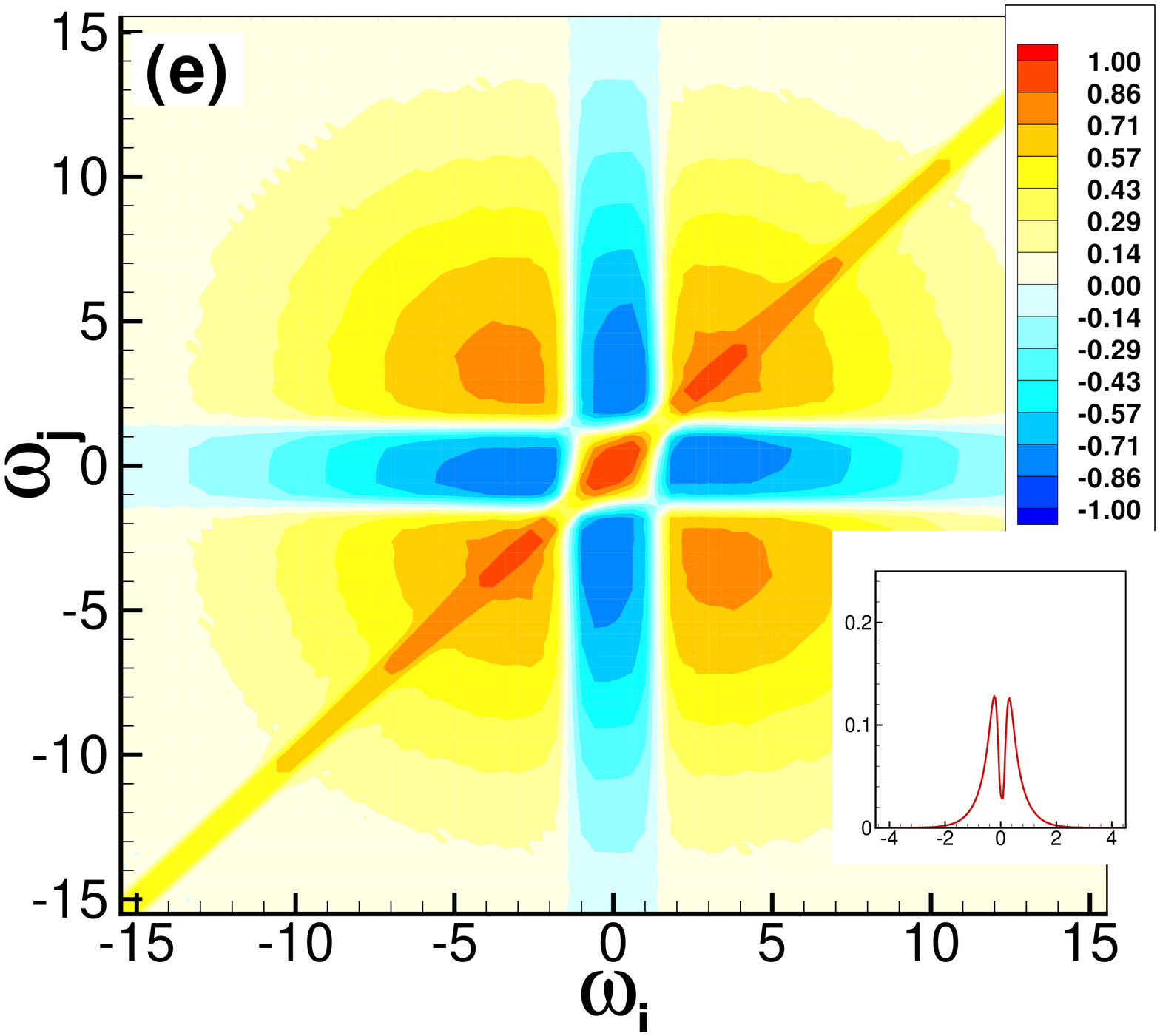}
\includegraphics[width=1.6in]{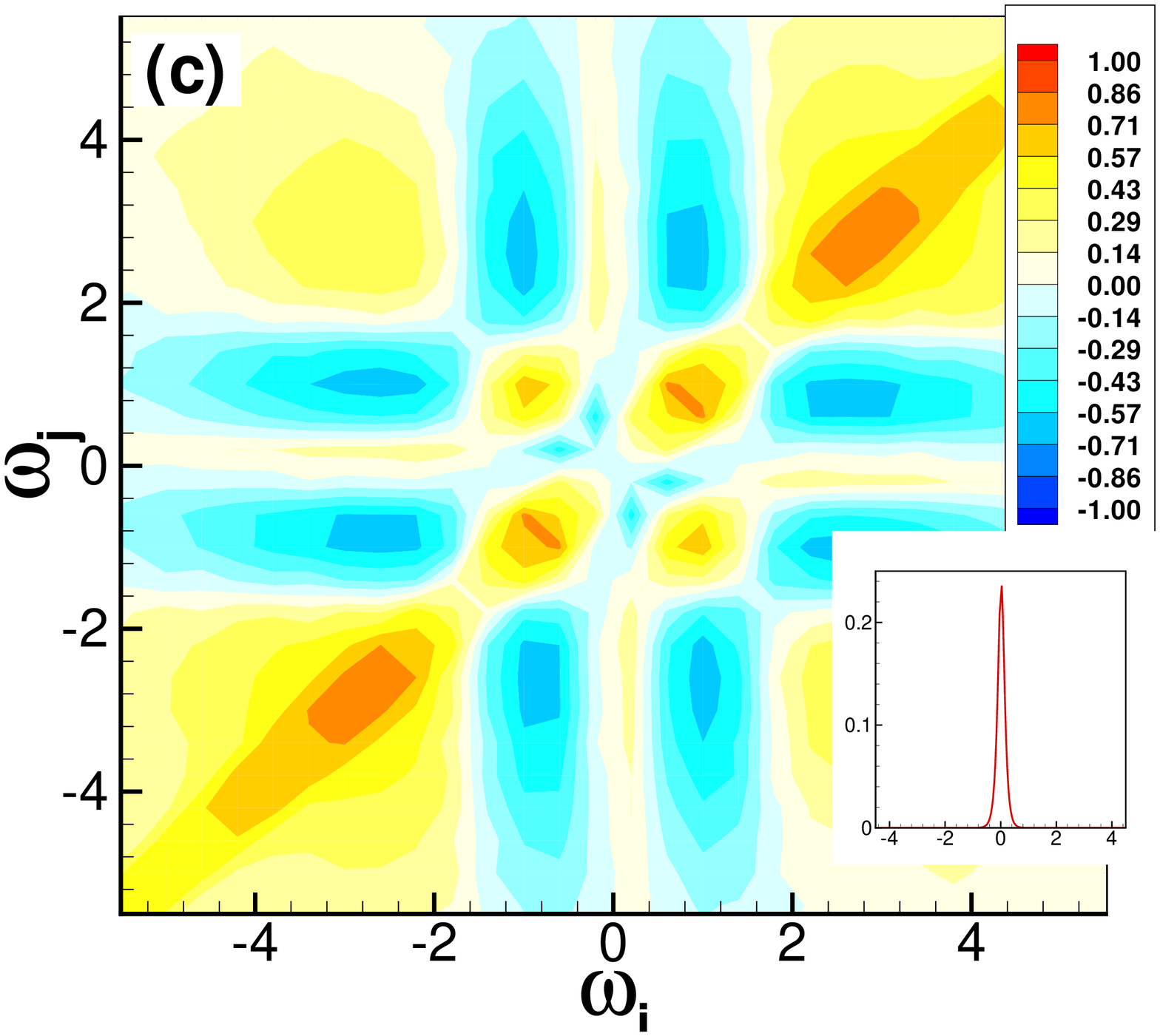}
\includegraphics[width=1.6in]{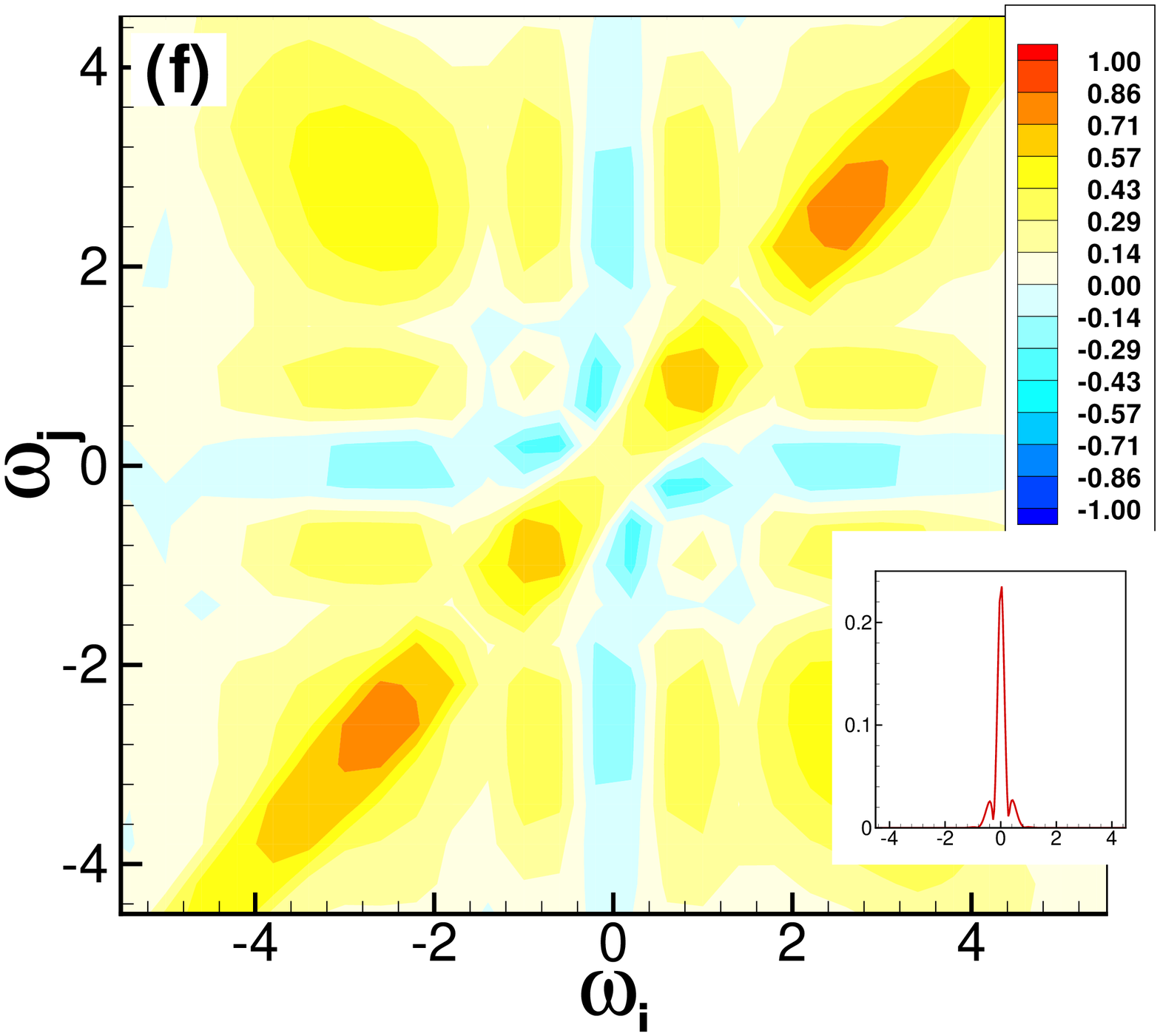}
\caption{Correlation spectra in the frequency domain for $N = 2$ soliton bound states. (a)-(c) an initial single-hump soliton, $\eta_1:\eta_2=1:3$, at different propagation distances, $3.1$, $3.9$, and $4.7$, respectively. (d)-(f) an initial double-hump soliton, $\eta_1:\eta_2=1:2$, at different propagation distances, $2.3$, $3.5$, and $4.7$, respectively. Insets show the soliton Fourier components. In all cases, $\eta_1+\eta=2$.}
\label{fig_f3}
\end{figure}
\end{center}

In addition, in Fig. \ref{fig_f3} we present the results of our calculations of the frequency-domain photon-number correlation spectra for the $N=2$ solitons with two values, $\eta_1:\eta_2 = 1:3$ and $\eta_1:\eta_2 = 1:2$.
The correlation coefficients, which are defined through the normally ordered covariance,
\begin{equation}
C_{ij}\equiv \frac{\langle :\Delta \hat{n}_{i}\Delta\hat{n}_{j}:\rangle }{\sqrt{\Delta \hat{n}_{i}^{2}\Delta
\hat{n}_{j}^{2}}}~,
\label{C}
\end{equation}
are calculated by means of the back-propagation method \cite{Lai95}. In Eq. (\ref{C}), $\Delta \hat{n}_{j}$ is the
photon-number fluctuation in the $i$-th slot $\Delta \omega_{i}$ in the frequency domain,
\[
\Delta \hat{n}_{i}=\int_{\Delta \omega_{i}}d\,t[U(z,\omega)\Delta \hat{U}^{\dag}(z,\omega)+U^{\ast }(z,\omega)\Delta \hat{U}(z,\omega)],
\]
where $\Delta \hat{U}(z,\omega)$ is the perturbation of the quantum-field operator, $U(z,\omega)$ is the classical unperturbed solution, and the integral is taken over the given spectral slot.

First, we reproduce the results for the photon-number correlation spectra of the initial $sech$-like single-hump solitons at $\eta_1:\eta_2 = 1:3$, reported earlier by Schmidt \textit{et al}.~\cite{Schmidt00, Schmidt-oc}.
A cross-pattern of the anti-correlated components, corresponding to the values $C_{ij}=-1$, occurs when the solitons merge, as shown in Fig.~\ref{fig_f3}(b).
This is the reason why an efficient number squeezing of the NLSE solitons can be produced by spectral filtering that removes the noisy spectral components \cite{Friberg}.
In between, the photon-number correlation spectra change periodically as the classical soliton profiles Fig.~\ref{fig_f3}(a-c)
It must be noted that there are some correlated patterns outside the center part of the solitons, even though the amplitude of the Fourier components there almost vanishing.
These correlated components in the far fringes come from the breathering dynamics of the $N=2$ solitons.

Now we turn to the case of a initially double-hump profile, $\eta_1:\eta_2=1:2$, as presented in Figs.~\ref{fig_f3}(d-f).
As can be seen in Fig.~\ref{fig_f3}(b) and Fig.~\ref{fig_f3}(e), the spectra for a double-hump pulse contains the same correlation patterns as that for a single-hump soliton when all of them merge, at $z = 3.9$, for $\eta_1/\eta_2 = 1:3$, and $z = 3.7$, for $\eta_1/\eta_2 = 1:2$.
That is what one expects for both of them having the same energy and same profile when the two solitons merge.
And again, the correlation spectrum for a double-hump soliton returns to the same pattern after a collision period, as shown in Fig.~\ref{fig_f3}(d) and Fig.~\ref{fig_f3}(f).
For double-hump solitons, there are significant differences in their correlation spectra patterns, see, e.g.,  Fig.~\ref{fig_f3}(a) and Fig.~\ref{fig_f3}(d).
If we only look at the center part of the correlation spectra where the soliton Fourier components are dominated, we can clearly see that there are strongly anti-correlated patterns for a double-hump soliton than in the case of a single-hump soliton.
This may be the reason that makes a double-hump $N=2$ soliton states get more squeezed than a single-hump one, although there are also some strongly anti-correlated patterns for a single-hump soliton in the far range where the soliton Fourier component almost vanish.
And only strongly {\em positive} correlated patterns occur in the center parts of both solitons, when the $N=2$ solitons merge into a single pulse.
Consequently, the optimal squeezing ratio of $N=2$ solitons degrades.

\section{Conclusions}
We have demonstrated that a two-soliton bound state gets more squeezed when it has a double-hump initial profile, and this effect is associated with longer soliton collision period.
We also study the photon-number correlation spectra of the $N=2$ solitons, which reveal the anti-correlated patterns which make the soliton to get more squeezed.
Since such double-hump two-soliton states have been generated in experiment, we do expect that our theoretical predictions can be readily verified experimentally, by generating strongly squeezed states for quantum information process and quantum computation.

We thank B. A. Malomed and E. A. Ostrovskaya for useful discussions and suggestions.


\begin{thebibliography}{99}
\bibitem{Ou92} Z. C. Ou, S. F. Pereira, H. J. Kimble, and K. C. Peng,  \textit{Phys. Rev. Lett.}
\textbf{68}, 3663 (1992).

\bibitem{Furusawa98} A. Furusawa, J. L. S{\o }rensen, S. L. Braunstein, C.
A. Fuchs, H. J. Kimble, and E. S. Polzik, \textit{Science} \textbf{282}, 706 (1998).

\bibitem{Silberhorn01} Ch. Silberhorn, P. K. Lam, O. Weib, F. K{\"{o}}nig,
N. Korolkova, and G. Leuchs, \textit{Phys. Rev. Lett.} \textbf{86}, 4267 (2001).

\bibitem{Silberhorn02} Ch. Silberhorn, N. Korolkova, and G. Leuchs, \textit{Phys. Rev. Lett.}
\textbf{88}, 167902 (2002).

\bibitem{Glockl03} O. Gl{\"{o}}ckl, S. Lorenz, C. Marquardt, J. Heersink, M.
Brownnutt, C. Silberhorn, Q. Pan, P. van Loock, N. Korolkova, and
G. Leuchs,  \textit{Phys. Rev. A} \textbf{68}, 012319 (2003).

\bibitem{Konig02} F. K{\"{o}}nig, M. A. Zielonka, and A. Sizmann,
\textit{Phys. Rev. A} \textbf{66}, 013812 (2002).

\bibitem{Carter87} S. J. Carter, P. D. Drummond, M. D. Reid, and R. M.
Shelby, \textit{Phys. Rev. Lett.} \textbf{58}, 1841 (1987).

\bibitem{Drummond87}P. D. Drummond and S. J. Carter, \textit{J. Opt. Soc. Am. B} \textbf{4}, 1565 (1987).

\bibitem{Lai89a}
Y. Lai and H. A. Haus,
{\it Phys. Rev. A.} \textbf{40}, 844 (1989); \textit{ibid} \textbf{40}, 854 (1989).

\bibitem{Lai89b}
Y. Lai and H. A. Haus,
{\it Phys. Rev. A.} \textbf{40}, 854 (1989).

\bibitem{Lai90} Y. Lai and H. A. Haus, \textit{Phys. Rev. A} \textbf{42}, 2925 (1990).

\bibitem{Friberg} S. R. Friberg, S. Machida, M. J. Werner, A. Levanon, and T. Mukai, \textit{Phys. Rev. Lett.} \textbf{77}, 3775 (1996).

\bibitem{RK-fbg} R.-K. Lee and Y. Lai, \textit{Phys. Rev. A} \textbf{69}, 021801(R) (2004).

\bibitem{Schmidt00} E. Schmidt, L. Kn{\" o}ll, D. Welsch, M. Zielonka, F. K{\" o}nig, and A. Sizmann, \textit{Phys. Rev. Lett.} \textbf{85}, 3801 (2000).

\bibitem{Lai95} Y. Lai and S.-S. Yu, \textit{Phys. Rev. A} \textbf{51}, 817 (1995).

\bibitem{Schmidt99} E. Schmidt, L. Kn{\" o}ll, and D.-G. Welsch, textit{Phys. Rev. A} {\bf 59}, 2442 (1999).

\bibitem{Rosenbluh}M. Rosenbluh and R. Shelby, \textit{Phys. Rev. Lett.} \textbf{66}, 153 (1991).

\bibitem{Yu01} C. X. Yu, H. A. Haus, and E. P. Ippen, \textit{Opt. Lett.} \textbf{26}, 669 (2001).

\bibitem{Werner97}M. J. Werner and S. R. Friberg, \textit{Phys. Rev. Lett.} \textbf{79}, 4143 (1997).

\bibitem{Zakharov}V. E. Zakharov and A. B. Shabat,
\textit{Sov. Phys. JETP} \textbf{34}, 62 (1972).

\bibitem{Werner96}M. J. Werner, \textit{Phys. Rev. A} \textbf{54}, 2567 (1996).

\bibitem{Yeang99}C.-P. Yeang,
\textit{J. Opt. Soc. Am. B} \textbf{16}, 1269 (1999).


\bibitem{Schmidt-oc}E. Schmidt, L. Kn{\" o}ll, D.-G. Welsch,
\textit{Opt. Commun.} \textbf{179}, 603 (2000).

\bibitem{s-laser}
H. A. Haus and M. N. Islam,
\textit{IEEE J. Quantum. Electron.} \textbf{21}, 1172 (1985).

\bibitem{surface-mode}
V. I. Gorentsveig, Yu. S. Kivshar, A. M. Kosevich, and E. S. Syrkin,
\textit{Int. J. Engng Sci.} \textbf{29}, 271 (1991).

\bibitem{Haus90} H. A. Haus and Y. Lai,
\textit{J. Opt. Soc. Am. B} \textbf{7}, 386 (1990).

\bibitem{Lai93} Y. Lai, \textit{J. Opt. Soc. Am. B} \textbf{10}, 475 (1993).


\end{thebibliography}
\end{document}